\documentclass[aps,prd,twocolumn,superscriptaddress,altaffilletter,nobibnotes,nofootinbib,10pt,showpacs,showkeys,preprintnumbers]{revtex4-2}

\usepackage{float}
\usepackage{graphicx}
\usepackage{dcolumn}
\usepackage[dvipsnames]{xcolor}
\usepackage{bm}
\usepackage{lineno}
\usepackage{float}
\usepackage{amsmath}
\usepackage{soul}
\usepackage{diffcoeff} 
\RequirePackage{multirow}

\usepackage{tikz,xcolor,hyperref}

\definecolor{lime}{HTML}{A6CE39}
\DeclareRobustCommand{\orcidicon}{
	\begin{tikzpicture}
	\draw[lime, fill=lime] (0,0) 
	circle [radius=0.16] 
	node[white] {{\fontfamily{qag}\selectfont \tiny ID}};
	\draw[white, fill=white] (-0.0625,0.095) 
	circle [radius=0.007];
	\end{tikzpicture}
	\hspace{-2mm}
}
\foreach \x in {A, ..., Z}{\expandafter\xdef\csname orcid\x\endcsname{\noexpand\href{https://orcid.org/\csname orcidauthor\x\endcsname}
			{\noexpand\orcidicon}}
}

\newcommand{\pte}{$p_{\rm{_T}}$}
\newcommand{\pts}{$p_{\rm{_T}}$~}

\newcommand{\pb}{$Pb-Pb$~}

\usepackage{soul}
\usepackage{float}

\definecolor{lime}{HTML}{A6CE39}
\DeclareRobustCommand{\orcidicon}{
	\begin{tikzpicture}
	\draw[lime, fill=lime] (0,0) 
	circle [radius=0.16] 
	node[white] {{\fontfamily{qag}\selectfont \tiny ID}};
	\draw[white, fill=white] (-0.0625,0.095) 
	circle [radius=0.007];
	\end{tikzpicture}
	\hspace{-2mm}
}
\foreach \x in {A, ..., Z}{\expandafter\xdef\csname orcid\x\endcsname{\noexpand\href{https://orcid.org/\csname orcidauthor\x\endcsname}
			{\noexpand\orcidicon}}
}

\begin{document}

\title{Study of Isothermal Compressibility and Speed of Sound in the Hadronic Matter Formed in Heavy-Ion Collision using Unified Formalism}

\author{Shubhangi Jain}%
\affiliation{Department of Physical Sciences, Indian Institute of Science Education and Research (IISER) Mohali, Sector 81 SAS Nagar, Manauli PO 140306 Punjab, India}
\author{Rohit Gupta\orcidA{}}
\affiliation{Shaheed Mangal Pandey Government Girls Degree College (SMPGGDC), Jananayak Chandrasekhar University (JNCU), Ballia 277001, Uttar Pradesh, India}

\author{Satyajit Jena\orcidB{}}
\email{sjena@iisermohali.ac.in}
\affiliation{Department of Physical Sciences, Indian Institute of Science Education and Research (IISER) Mohali, Sector 81 SAS Nagar, Manauli PO 140306 Punjab, India}

\begin{abstract}
The thermodynamical quantities and response functions are useful to describe the particle production in heavy-ion collisions as they reveal crucial information about the produced system. While the study of isothermal compressibility provides an inference about the viscosity of the medium, speed of sound helps in understanding the equation of state. With an aim towards understanding the system produced in the heavy-ion collision, we have made an attempt to study isothermal compressibility and speed of sound as function of charged particle multiplicity in heavy-ion collisions at $\sqrt{s_{NN}}$ = $2.76$ TeV, $5.02$ TeV, and $5.44$ TeV using unified formalism.
\end{abstract}

\maketitle

\section{Introduction}
Among the main aim of heavy-ion collision program at present collider experiments such as Relativistic Heavy Ion Collider (RHIC) and Large Hadron Collider (LHC) is to mimic the state that was created few microseconds after the Big Bang. This state of matter, created at extremely high temperature and energy density, is called the Quark-Gluon Plasma (QGP) and is also believed to be present at the core of massive neutron stars. Interaction between quark and gluon, which leads to the formation of QGP, is governed by the Quantum Chromo-dynamics (QCD). Asymptotic freedom, which is an important pillar of QCD, suggests a confinement deconfinement phase transition, during~which the hadronic degree of freedom changes to partonic degree of freedom. Whether the phase transition is first-order, second-order or a simple cross-over and the search for the critical point are some of the important questions that are of immediate interest in the particle physics~community. 

Since the formation of QGP occurs at a very short time scale, it is not possible to directly probe in the experiment using current technologies. Therefore, we rely on the information carried by the final state particles to the detectors to gain insight into the medium created in the heavy-ion collision. Although~we only measure the kinematic quantities such as the pseudorapidity $\eta$, transverse momentum \pte, energy $E$ etc.~of the final state particles in the experiment, a~breadth of information about the medium can be extracted by studying these kinematic~observables. 

Another set of quantities that are not directly observable, however, play an important role in understanding the nature of the medium and the equation of state are the thermodynamical response functions. This includes quantities that express how a system responds to change in some external parameters such as pressure, temperature etc. Isothermal compressibility ($\kappa_T$), and~speed of sound ($c_s$) are some of the response function that are of interest in high-energy physics~\cite{Sahu:2020swd, Basu:2016ibk, Khuntia:2016ikm}. The~isothermal compressibility, $\kappa_{T}$, which exhibits the important property of the medium, tells us how much the volume of the medium changes on the change in pressure at a fixed temperature. This quantity can be used to study how close a medium is to be called perfect fluid. Perfect fluids are ideal fluids that do not possess shear stress, viscosity and also do not conduct heat. The~$\kappa_{T}$ of perfect fluid is zero and the zero value signifies that the fluid is incompressible. Although~the incompressible fluids do not exist in nature, the~recent findings of the value of $\kappa_{T}$, as~in Ref.~\cite{Sahu:2020nbu}, are almost close to zero which suggests that the medium created is almost a perfect fluid. Estimation of $\kappa_{T}$ will tell us about how close the medium to be a perfect fluid which will lead us an approximation about the viscosity as viscosity describes a fluid's resistance to flow, hence perfect fluid has zero viscosity. Isothermal compressibility and viscosity of a liquid is related to the low-shear viscosity of a liquid dispersion of solid particles~\cite{Mezzasalma2002AnEF}. Perfect fluid can also be characterized by the ratio of shear viscosity to entropy density ($\eta$/$s$). Calculation based on $AdS/CFT$ correspondence has put up a universal lower bound of $1/4\pi$ for strongly interacting quantum field theories~\cite{Kovtun:2004de}. On~the other hand,  the~value of $\eta$/$s$ has been found to be close to the lower bound based on the flow harmonics calculation of the experimental data, indicating the near-perfect behaviour of medium created in heavy-ion collision~\cite{ALICE:2011ab, Luzum:2008cw}. 

As explained in Ref.~\cite{Bjorken:1982qr}, the~speed of sound can quantify the nature of the same as it connects and explains the hydrodynamical evolution of the produced matter in the heavy-ion collisions. Fundamentally the speed of sound also gives the information about the equation of state, which relates pressure $(P)$ and the energy density $(\epsilon)$. For~a non-interacting massless ideal gas, the~value of the squared speed of sound $c_s^2$ is expected to be $1/3$ times speed of light squared~\cite{Hallman:2002qi}. Hence, the~comparison with the massless ideal gas will give crucial information about the system dynamics and reveals the nature of the medium~\cite{Deb:2019yjo}. Different studies suggested that for the system created in heavy-ion collision the value of $c_s^2$ is close to the ideal value~\cite{Sahu:2020swd, Deb:2019yjo, Deb:2020ezw, Tiwari:2017aon, Khuntia:2016ikm, Castorina:2009de, NasserTawfik:2012jpa, Deppman:2012qt, Gardim:2019xjs}.

As already discussed, these quantities are not directly observable in the experiment and we extract them by utilizing the distribution of kinematic observables such as the transverse momentum \pte-spectra, rapidity, angle of emission etc. The~\pte-spectra carries sufficient information to study such quantities as it is directly related to the energy of the system. Understanding the distribution of \pts is in itself a tedious task because in the low-\pts region, the~QCD coupling strength is very high and hence we cannot apply the perturbative QCD theories to explain the spectra. Several phenomenological models have been developed to tackle this issue, and~the most widely accepted are the statistical thermal models. We can utilize the statistical thermal models to extract the thermodynamical quantities such as temperature, number density, energy density~etc.

If we assume the purely thermal origin of final state particles, the~most natural choice to explain the energy distribution of particles is Boltzmann-Gibbs (BG) statistics~\cite{Schnedermann:1993ws, Stodolsky:1995ds, Sharma:2018jqf}.  However, it has been discussed in many works~\cite{Jena:2020wno, Gupta:2020naz} that the BG distribution function deviates significantly from the experimental data because the spectra are more like a power-law rather than the simple exponential. Also, the~BG statistics fails to explain the strongly correlated systems~\cite{TSALLIS1995539} in which the long-range correlations are present, and~entropy becomes non-additive and non-extensive~\cite{lemanska2012nonadditive}. The~existence of long-range interaction in high-energy heavy-ion collisions is discussed in Ref.~\cite{Alberico:1999nh} motivating to explore beyond the extensive BG regime to study the spectra. In~1988, C. Tsallis proposed a \mbox{statistics~\cite{Tsallis:1987eu, Biro:2020kve, Parvan:2015asa}}, introducing an additional parameter $q$, which takes care of the non-extensivity in the system.  It is a thermodynamically consistent~\cite{Cleymans:2011in, Conroy:2010wt}, generalized version of Boltzmann distribution~\cite{Tsallis:1998ws}. The~power-law behavior of Tsallis distribution makes it a good choice to study the  \pte-spectra and it is shown to nicely fit the spectra, particularly in the low-\pts region. Although~Tsallis statistics nicely explain the data in the low-\pts region, however, it starts to deviate from the experimental data as we move toward the high-\pts part of the~spectra.

Particle spectra in heavy-ion collisions can be divided into two distinct regions, low-\pts regime corresponds to the particle produced in soft processes whereas the hard processes dominate particle production in the high-\pts region. The~limitation of Tsallis statistics in explaining the particle produced in hard processes demands a framework that can consider the effect of both soft and hard processes in the particle spectra. Some modification in Tsallis statistics~\cite{Azmi:2015xqa, Cirto:2014sra, Wong:2013sca, Wong:2014uda} has been proposed to explain the high-\pts part of spectra in the heavy-ion collision, however, more work is required in this direction to get the full benefit from the spectra. To~explain both the hard and soft part of particle spectra in a consistent manner, a~unified theory using Pearson distribution is introduced in Ref.~\cite{Jena:2020wno}. It is a generalized form of the Tsallis distribution and is shown to be thermodynamically consistent and backward compatible to the Tsallis statistics within some limit on its parameters~\cite{Gupta:2020naz}. 
 
In this work, we have calculated the isothermal compressibility and speed of sound for charged hadrons produced in heavy-ion collisions using the unified statistical framework. For~this analysis, we have taken the experimental data of transverse momentum spectra for charged hadrons produced in \pb collision at $\sqrt{s_{NN}}$ = $2.76$ TeV~\cite{Abelev:2012hxa}, $5.02$ TeV~\cite{ALICE:2018vuu}, and~$Xe-Xe$ collision at $5.44$ TeV~\cite{ALICE:2018hza} measured by the ALICE~experiment.

\section{Methodology} \label{sec:methodology}
The basic thermodynamic quantities that are of interest to formulate the isothermal compressibility and speed of sound include energy density $\epsilon$, number density $n$ and pressure $P$. From~the standard thermodynamics, the~number of particles $N$ in a system and its total energy $E$ can be calculated as:
\begin{equation} \label{eq1}
    N = \sum_{i} f_{i}
\end{equation}
\begin{equation}  \label{eq2}
    E = \sum_{i} E_{i} f_{i}
\end{equation}
where $E_{i}$ is the energy of $i^{th}$ state and $f_{i}$ is the corresponding distribution function. The~standard replacement while going from summation to integration for small energy intervals is given as~\cite{Cleymans:2011in}:
\begin{equation} \label{eq3}
    \sum_{i} \to V \int \frac{d^{3}p}{(2 \pi)^{3}}
\end{equation}

Here, $V$ is the volume and $p$ represent the momentum. So, using the above transformation, the~number density $n$ will be of the form:
\begin{equation} \label{eq4}
    n = \int \frac{d^{3}p}{(2\pi)^{3}} \times f(E)
\end{equation}
and the corresponding energy density $\epsilon$ will be given as:
\begin{equation} \label{eq5}
\epsilon = \int \frac{d^{3}p}{(2\pi)^{3}} E \times f(E)
\end{equation} 

Since the momentum distribution of the final state particles are fixed at kinetic \mbox{freeze-out~\cite{Deb:2020ezw}}, the~pressure of the system could be estimated from the moments of energy distribution.
The pressure $P$ is given as:
\begin{equation} \label{eq6}
P = \int \frac{d^{3}p}{(2\pi)^{3}} \frac{p^{2}}{3E} \times f(E)
\end{equation}

Among all the quantities discussed above, one common factor is the energy distribution of the particles ($f(E)$). Energy is related to the transverse mass $m_T$ and pseudorapadity $y$ as $E = m_T cosh(y)$ and the transverse mass is defined in term of transverse momentum $p_T$ and mass of particle $m$ as $m_T = \sqrt{p_T^2 + m^2}$.  So, the~distribution of transverse momenta acts as a proxy for the energy distribution. Hence, the~proper parameterization of transverse momentum spectra is crucial to understand the thermodynamics of the system created in high-energy collisions. In~the present work, we have used the unified statistical framework to explain the \pte-spectra and extract the thermodynamical quantities such as temperature $T$, non-extensive parameter $q$. 
 
In the seminal work~\cite{Pearson343}, Karl Pearson discussed a family of the curve, based on the first four moments (mean, variance, skewness, and~kurtosis), called Pearson distribution. Before~the introduction of Pearson formalism in 1895, all probability distribution was only constructed based on mean and variance and did not take care of skewness and kurtosis. Pearson introduced a new probability distribution function where skewness and kurtosis can also be adjusted along with the mean and variance of a distribution. An~important characteristic of this distribution is that depending on the limit on its parameters, it reduces to different distribution function such as Gaussian, normal, Student's T, Gamma distribution etc. The~differential form of a Pearson distribution function, $p(x)$, for~a variable $x$ is expressed as~\cite{pollard}:
\begin{equation}\label{eq7}
 \frac{1}{p(x)}\frac{dp(x)}{dx} + \frac{a + x}{b_0 + b_1 x + b_2 x^2} = 0
\end{equation}
where a, $b_{0}$, $b_{1}$, and~$b_{2}$ are related to first four moments of the distribution. By~integrating this differential equation, one can get,
\begin{equation}  \label{eq8}
    p(x) = \exp \Bigg(- \int \frac{x+a}{b_{2}x^{2}+b_{1}x+b_{0}}dx\bigg)
\end{equation}

 Solving above equation we get the general solution of the form:
\begin{equation}
     p(x) = C(e+x)^f(g+x)^h
 \end{equation}
\begin{equation}  \label{eq9}
    p(x) = B\bigg(1+\frac{x}{e}\bigg)^{f}\bigg(1+\frac{x}{g}\bigg)^{h}
\end{equation}
upto some normalization constant $B = Ce^fg^h$. Here, $C,~e,~f,~g~\&~h$ are the parameters of the equation and can be related to the physical parameters such as temperature $T$, non-extensivity parameter $q$ etc.

Distribution function, in~case of unified statistical framework, obtained from the above Pearson distribution, is given as~\cite{Gupta:2020naz}:
\begin{equation}  \label{eq10}
f_{i} = (Bf_{E})^{1/q} f_{Ta}
\end{equation}
where
\begin{equation}  \label{eq11}
B = \frac{C}{(p_{0})^{n}} \bigg(\frac{T}{q-1}\bigg)^{\frac{-q}{q-1}} 
\end{equation}
\begin{equation}  \label{eq12}
f_{E} = \frac{1}{E} \bigg(1+\frac{E}{p_{0}}\bigg)^{-n}
\end{equation}
and
\begin{equation}  \label{eq13}
f_{Ta} = \bigg[1+(q-1)\frac{p_{T}}{T}\bigg]^{\frac{-1}{q-1}}
\end{equation}

In the above equation, $p_0$ and $n$ are the free parameters with parameter $n$ is related to the second order flow coefficient~\cite{Jena:2020wno}. This formalism reduces to Tsallis statistics within the limit $n=-1$ and $p_{0} =0$.
 Therefore, it can be considered as a generalized version of the Tsallis function and explains both soft and hard process contributions to \pte-spectra. The~equation for the average number of particles and energy, in~the case of unified formalism, remains the same as Tsallis~\cite{Gupta:2020naz}:
\begin{equation}  \label{eq14}
N = \sum_{i} f_{i}^{q}
\end{equation}
and, the~energy of the system will be:
\begin{equation}  \label{eq15}
E = \sum_{i} E_{i} f_{i}^{q}
\end{equation}

Here, the~additional power of $q$ comes from the thermodynamic consistency. In~case of the unified formalism, the~transverse momentum spectra is defined as:
\begin{equation}  \label{eq16}
     \frac{1}{2\pi p_{T}} \frac{d^{2}N}{dp_{T}dy} = B^\prime \bigg(1+\frac{p_{T}}{p_{0}}\bigg)^{-n}
     \bigg[1+(q-1)\frac{p_{T}}{T}\bigg]^{\frac{-q}{q-1}}
\end{equation}
where $B^\prime = B \times \frac{V}{(2\pi)^{3}}$, $T$ is temperature and $q$ is non-extensive parameter. Here we considered the chemical potential to be zero because at LHC energy, the~net-baryonic number is extremely small at the central rapidity region. Thermodynamic parameters such as $T$, $q$ and the other quantities can be obtained by fitting the measured transverse momentum spectra with the unified distribution using the Equation~(\ref{eq16}). These quantities extracted from the spectra can be used to calculate the response function as discussed~below.

\subsection{Isothermal~Compressibility}
In the high-energy collider experiment, we only consider a part of phase space because of the limited $\eta$ acceptance of the detectors. The~overall number of particles in the collision is conserved, but~number of particles and energy in a particular phase space window may vary. Hence, the~system can be considered as a grand canonical ensemble for the estimation of isothermal compressibility. Therefore, the~variance of number of particles N, can be written as~\cite{kardar_200}:
\begin{equation}\label{eq:kT_one}
    \big \langle(N - \langle N \rangle)^2 \big \rangle = VT\frac{\partial n}{\partial \mu}
\end{equation}

And, the~isothermal compressibility, $\kappa_T$, can be written as:
\begin{equation} \label{eq:kT_standard}
    \kappa_{T} = -\frac{1}{V} \diffp{V}P{T}
\end{equation}

Using the expression of variance of $N$ and  $\kappa_T$ \cite{kardar_200, Mrowczynski:1997kz}, we can write:
\begin{equation}\label{eq:kT_two}
    \big \langle(N - \langle N \rangle)^2 \big \rangle = var(N) = TV n^2 \kappa_T
\end{equation}

Equation~(\ref{eq:kT_two}) requires an event-by-event information of N to estimation $\kappa_T$. On~contrary, we can compare Equations~(\ref{eq:kT_one}) and (\ref{eq:kT_two}) to derive a fluctuation independent formula for isothermal compressibility as:
\begin{equation}\label{eq:kT_final}
    \kappa_{T} =\frac{\partial n/\partial \mu}{n^2} 
\end{equation}

As per Equation~(\ref{eq:kT_final}), the~estimation of $\kappa_T$ does not depend on the fluctuation of N and it  makes the estimation possible without having event-by-event information of particle~number.

Number density, $n$, in~case of unified formalism, is of the form:
\begin{equation} \label{eq21}
    n = \int \frac{d^{3}p}{(2\pi)^{3}} \times \frac{B}{E}\bigg(1+\frac{E}{p_{0}}\bigg)^{-n}\bigg[1+(q-1)\frac{(E-\mu)}{T}\bigg]^{\frac{-q}{q-1}}
\end{equation}
and,
\begin{equation}  \label{eq22}
  \diffp{n}\mu = \int \frac{d^{3}p}{(2\pi)^{3}} \times \frac{q}{T} \times \frac{B}{E}\bigg(1+\frac{E}{p_{0}}\bigg)^{-n}\bigg[1+(q-1)\frac{(E-\mu)}{T}\bigg]^{\frac{1-2q}{q-1}} 
\end{equation}

By using the above equations, we have estimated the values of $\kappa_{T}$ for heavy-ion collisions at different~energies.

\subsection{Speed of~Sound}
For a thermodynamic system at temperature $T$ and volume $V$, the~squared speed of sound is given by,
\begin{equation} \label{eq23}
c_{s}^{2} = \diffp{P}\epsilon{s}
\end{equation}
where $P$ is pressure and $\epsilon$ is energy density of the system. As~discussed in Ref.~\cite{LANDAU1987251}, the~propagation of sound wave in a medium is an adiabatic process and entropy is constant in such process, hence the squared speed of sound is estimated at constant entropy \mbox{density ($s$)}. Above~equation can be further reduced to:
\begin{equation} \label{eq24}
 c_{s}^{2} =\dfrac{\diffp{P}T}{\diffp{\epsilon}T} 
\end{equation}
where
\begin{equation}
    P = \int \frac{d^{3}p}{(2\pi)^{3}} \times B \times \frac{p^{2}}{3E^{2}}\bigg(1+\frac{E}{p_{0}}\bigg)^{-n}\bigg[1+(q-1)\frac{E}{T}\bigg]^{\frac{-q}{q-1}}
\end{equation}
and,
\begin{equation}
    \epsilon = \int \frac{d^{3}p}{(2\pi)^{3}} \times B \bigg(1+\frac{E}{p_{0}}\bigg)^{-n}\bigg[1+(q-1)\frac{E}{T}\bigg]^{\frac{-q}{q-1}}
\end{equation}

By using the above equations, the~squared speed of sound $c_{s}^{2}$ reduces to
\begin{equation}   \label{eq27}
    c_{s}^{2} = \frac{\int \frac{p^{2}d^{3}p}{3E^{2}} \bigg(1+\frac{E}{p_{0}}\bigg)^{-n} \bigg[\frac{T}{q-1}+E\bigg]^{\frac{1-2q}{q-1}}}{\int d^{3}p \bigg(1+\frac{E}{p_{0}}\bigg)^{-n} \bigg[\frac{T}{q-1}+E\bigg]^{\frac{1-2q}{q-1}}}
\end{equation}

We have used the Equation~(\ref{eq27}) to estimate the squared speed of sound in the medium created in heavy-ion collision at three different~energies.

\section{Results and~Discussion}  \label{sec:discussion}

This study presents a formalism to calculate $\kappa_{T}$ and $c_{s}^{2}$ using the non-extensive unified statistical framework discussed in Ref.~\cite{Gupta:2020naz}. We have estimated the $\kappa_{T}/V$ and $c_{s}^{2}$ in the medium formed of charged hadrons as a function of charged particle multiplicity for different collision systems. The~data for charged particle multiplicity $(\big<\diff{N_{ch}}\eta\big>)$ corresponding to a particular centrality is taken from the experimental results Refs.~\cite{Abelev:2013vea, Acharya:2019yoi, ALICE:2018cpu}.

For this analysis, we have considered the transverse momentum spectra of charged hadrons produced in $Pb-Pb~$ collision at $2.76$ \cite{Abelev:2012hxa} and $5.02$ TeV~\cite{ALICE:2018vuu} and $Xe-Xe$ collision at $5.44$ TeV~\cite{ALICE:2018hza}. The~$p_T~$ range is restricted to $p_T<5$ GeV/c since we are trying to study bulk properties and the majority of high $p_T$ particles are produced from hard processes. The~experimental data for the $p_T$-spectra for all the energies used in the paper belongs to the pseudorapidity range $|\eta|<0.8$. The~unified function fit to the $p_T$-spectra at $2.76$, $5.02$ and $5.44$ TeV is provided in the Refs
. \cite{Jena:2020wno, Gupta:2021efj, Gupta:2021oxf}. The~numerical value of $T$, $q$ and the other fitting parameters are calculated by fitting the measured transverse momentum spectra with the unified formalism as in Equation~(\ref{eq16}) and the best fit value of the parameters are provided in Table~\ref{tab:pbpb}.

  \begin{table*}[t]
\caption{\label{tab:pbpb}Numerical values of the fit parameters $T$ (GeV), $q$, $p_0$ (GeV/c) and $n$  obtained by fitting the experimental data of $p_T$-spectra fitted with the unified formalism Eq.~(\ref{eq16}). }
\centering
{\scriptsize \begin{tabular*}{\textwidth}{@{\extracolsep{\fill}}lllllllllllll@{}}
\hline
\multirow{2}{*}{Centrality} & \multicolumn{4}{c|}{\textbf{$2.76$ TeV}} &  
\multicolumn{4}{c|}{\textbf{$5.02$ TeV}} &
\multicolumn{4}{c}{\textbf{$5.44$ TeV}}

\\ 
\cline{2-13}
&T & q & $p_0$ & n &T  & q & $p_0$ & n &T  & q & $p_0$ & n \\
\hline
\multirow{2}{*}{0--5\%} & 0.393 & 1.048 & 0.105 &  0.749 & 0.407 & 1.048 & 0.0018 &  0.562& -&-&-&-\\
&$\pm 0.05 $ & $\pm 0.004 $ & $\pm 0.21 $ & $ \pm 0.36 $ & $\pm0.003 $ & $\pm 0.000 $ & $\pm 0.030 $ & $\pm 0.05 $ & & & & \\
\hline
\multirow{2}{*}{5--10\%} & 0.386 & 1.053 &  0.0877 & 0.700 & 0.415 & 1.049 & 0.0167 & 0.604 &-&-&-&-\\
&$\pm 0.04 $ & $\pm 0.041 $ & $\pm 0.191 $ & $ \pm 0.32 $ & $\pm0.004 $ & $\pm 0.000 $ & $\pm 0.033 $ & $\pm 0.05 $ &  & & & \\
\hline
\multirow{2}{*}{10--20\%} & 0.370 & 1.060 & 0.0600 & 0.619 & 0.422 & 1.052 & 0.0394 & 0.659 & 0.409 & 1.072 & 0.0977 & 0.720\\
&$\pm 0.07  $ & $\pm 0.006 $ & $\pm 0.18 $ & $ \pm 0.30 $ & $\pm 0.004 $ & $\pm 0.000 $ & $\pm 0.033 $ & $\pm 0.06 $ & $\pm 0.01 $ & $\pm0.001 $ & $\pm0.08 $ & $\pm0.15 $ \\
\hline
\multirow{2}{*}{20--30\%} & 0.351 & 1.070 & 0.0385 & 0.548 & 0.424 & 1.059 & 0.0812 & 0.744 & 0.460 & 1.067 & 0.225 & 1.101\\
&$\pm 0.08 $ & $\pm 0.008 $ & $\pm 0.18 $ & $ \pm 0.30 $ & $\pm 0.012 $ & $\pm 0.001 $ & $\pm 0.042 $ & $\pm 0.07 $ & $\pm0.03 $ & $\pm0.003 $ & $\pm 0.11 $ & $\pm 0.24 $ \\
\hline
\multirow{2}{*}{30--40\%}& 0.331 &  1.081 & 0.0256 & 0.489 & 0.412 & 1.068 & 0.0824 & 0.749 & 0.447 & 1.079 & 0.2286 & 1.112\\
&$\pm 0.07 $ & $\pm 0.008 $ & $\pm 0.20 $ & $ \pm 0.34 $ & $\pm 0.013 $ & $\pm 0.001 $ & $\pm 0.038 $ & $\pm 0.07 $ & $\pm0.04 $ & $\pm0.004 $ & $\pm0.10 $ & $\pm0.24 $ \\
\hline
\multirow{2}{*}{40--50\%} & 0.311 & 1.093 &  0.0341 & 0.474 & 0.369 & 1.085 & 0.05 & 0.614 & 0.455 & 1.091 & 0.2881 & 1.306\\
&$\pm 0.08 $ & $\pm 0.008 $ & $\pm 0.25 $ & $ \pm 0.46 $ & $\pm 0.018 $ & $\pm 0.002 $ & $\pm 0.042 $ & $\pm 0.08 $ & $\pm 0.05 $ & $\pm 0.005 $ & $\pm 0.13 $ & $\pm0.36 $ \\
\hline
\multirow{2}{*}{50--60\%} & 0.292 & 1.106 & 0.0457 & 0.468 & 0.34 & 1.101 & 0.0527 & 0.578 & 0.434 & 1.108 & 0.2904 & 1.317\\
&$\pm 0.08 $ & $\pm 0.008 $ & $\pm 0.32 $ & $ \pm 0.61 $ & $\pm 0.023 $ & $\pm0.002 $ & $\pm 0.051 $ & $\pm 0.11 $ & $\pm0.08 $ & $\pm0.008 $ & $\pm0.15 $ & $\pm0.51 $ \\
\hline
\multirow{2}{*}{60--70\%} & 0.273  & 1.121  & 0.0747 &  0.487 & 0.311 & 1.118 & 0.0658 & 0.557 & 0.357 & 1.123 & 0.1977 & 0.943\\
&$\pm 0.11 $ & $\pm 0.012 $ & $\pm 0.49 $ & $ \pm 1.03 $ & $\pm 0.025 $ & $\pm0.002 $ & $\pm0.071$ & $\pm0.17 $ & $\pm0.07 $ & $\pm 0.006 $ & $\pm 0.17 $ & $\pm 0.53 $ \\
\hline
\multirow{2}{*}{70--80\%} & - & - & - & - & 0.329 & 1.131 & 0.1565 & 0.855 & 0.338 & 1.139 & 0.2060 & 0.974\\
&  &  &  &  & $\pm0.034 $ & $\pm0.003 $ & $\pm0.094 $ & $\pm0.29 $ & $\pm0.09 $ & $\pm0.011 $ & $\pm0.26 $ & $\pm 0.87 $ \\
\hline
 \end{tabular*}}
\end{table*}

\begin{figure}[H] 
\hspace{-10pt}
\includegraphics[height=2in,width=3in]{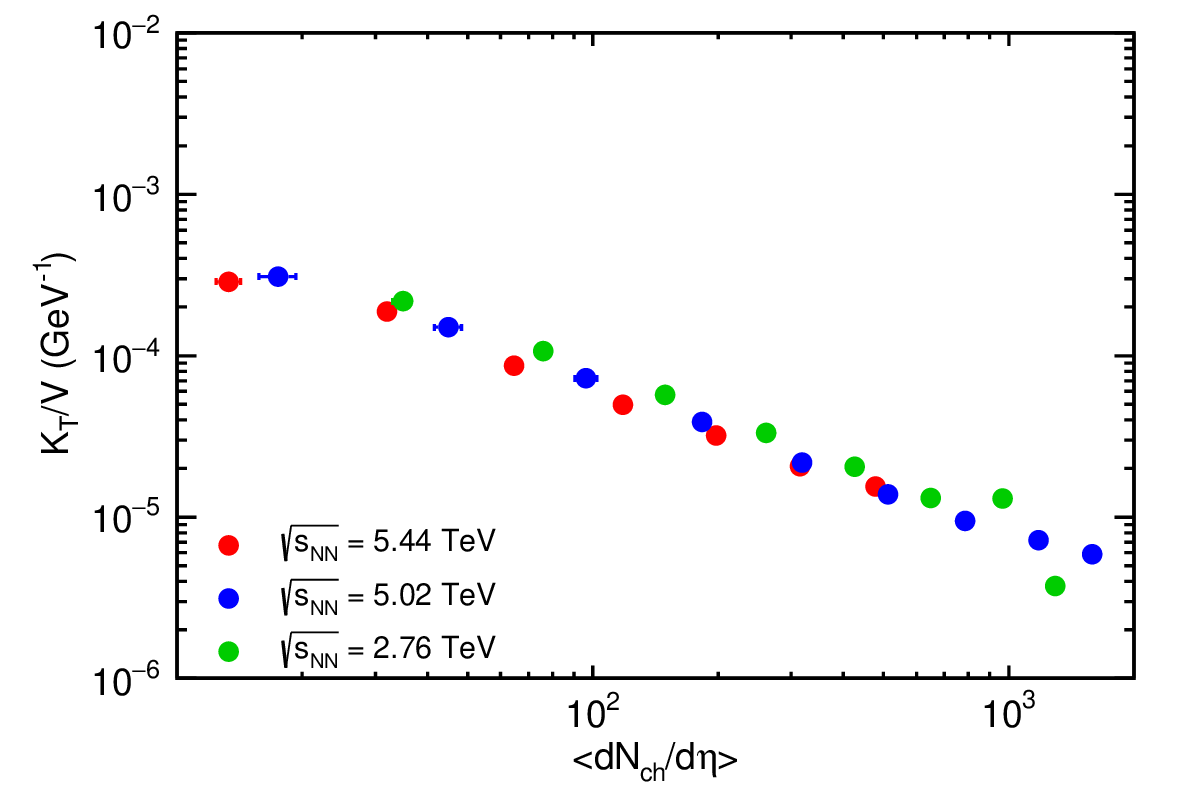}
\caption{ Variation of isothermal compressibility over volume $(\kappa_{T}/V)$ with the average charged particle multiplicity $(\big<\diff{N_{ch}}\eta\big>)$ for \pb collision at $\sqrt{s_{NN}}$ = $2.76$ TeV, \pb collision at $\sqrt{s_{NN}}$ = $5.02$ TeV and $Xe-Xe$ collision at $\sqrt{s_{NN}}$ = $5.44$ TeV using Unified formalism Eq.~(\ref{eq:kT_final}), Eq.~(\ref{eq21}) \& Eq.~(\ref{eq22})}
\label{Fig1}
\end{figure}

In Figure~\ref{Fig1}, we have plotted the isothermal compressibility over volume calculated using the Equations~(\ref{eq:kT_final})--(\ref{eq22}). It is observed that there is a decline in the values of $\kappa_{T}/V$ with an increase in the multiplicity.
At higher charged-particle multiplicity, $\kappa_{T}/V$ becomes the lowest, which suggests that the system move toward near-ideal behaviour with the increase in multiplicity. This trend is in line with the expectation as higher multiplicity class contains a larger number of particles and hence a higher pressure is required to attain a small change in volume. Similar values of $\kappa_{T}/V$ for different collision systems show an indication of similar dynamics of the produced medium. It is worth mentioning here that the ideal fluid is incompressible, hence $\kappa_T = 0$, implying that the volume cannot be changed by applying pressure. For~water, the~corresponding value is several order of magnitude higher than what is obtained in case of heavy-ion collision. The~values for $\kappa_T/V$ obtained in the case of heavy-ion collision using the unified formalism is in the range from  $~10^{-3}$ to $10^{-5}~ GeV^{-1}$.

A proper estimation of volume is required to extract the value of $\kappa_T~(fm^3/GeV)$. Different techniques have been developed and tested on diverse datasets to extract the volume parameter~\cite{Braun-Munzinger:2014lba, Cleymans:2012ya, Abelev:2014pja, Tawfik:2019oct, Gardim:2020sma, Azmi:2015xqa, Chatterjee:2015fua, BraunMunzinger:2003zz}. Although~the numerical values vary greatly in different models, all of them are in the order of $10^3-10^4~fm^3$ and hence utilizing the value of volume from these models will give us the value of $\kappa_T$ in the order of $1-10~fm^3/GeV$. This range of the value of $\kappa_T$ matches well with the values obtained by other techniques in the Ref.~\cite{Sahu:2020nbu, Khuntia:2018non}.
The obtained value of $\kappa_T$ is very low as compared to the water and other materials, indicating that the compressibility of the system created in the heavy-ion collision is very close to an ideal fluid. Proper estimation of volume is still an undergoing field of research, hence, we did not select a particular model and instead, we presented the value in terms of $\kappa_T/V$.

\begin{figure}[H] 
\hspace{-10pt}
\includegraphics[height=2in,width=3in]{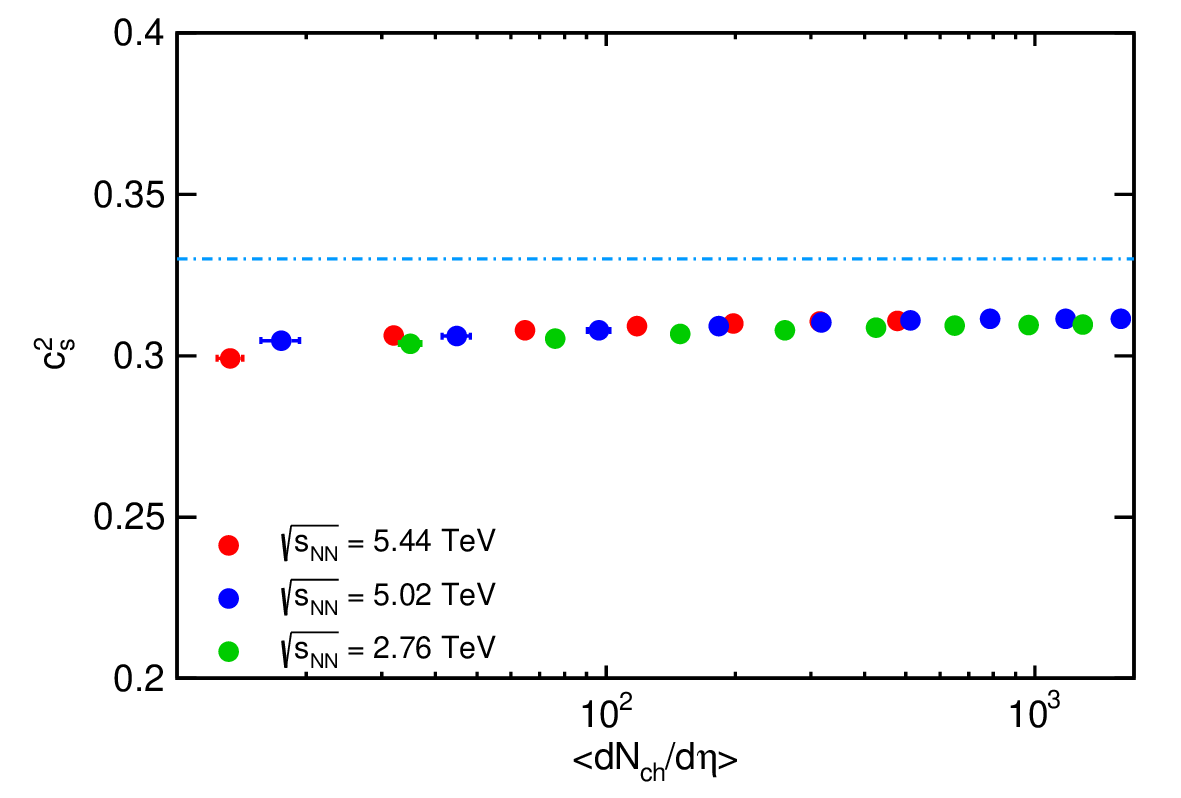}
\caption{Variation of squared speed of sound $(c_s^2)$ as a function of $\big<\diff{N_{ch}}\eta\big>$ for \pb collision at $\sqrt{s_{NN}}$ = $2.76$ TeV, \pb collision at $\sqrt{s_{NN}}$ = $5.02$ TeV and $Xe-Xe$ collision at $\sqrt{s_{NN}}$ = $5.44$ TeV using Unified formalism as Eq.~(\ref{eq27}). The dotted line represents the theoretical value for ideal gas system.}
\label{Fig2}
\end{figure}

 We have also attempted to study the speed of sound for different collision systems in order to explore the properties of matter. The~speed of sound in a medium reveals the properties of the medium via the equation of state. In~Figure~\ref{Fig2}, we have plotted the squared speed of sound with charged-particle multiplicity for three different energies estimated using the Equation~(\ref{eq27}). It is observed that the value of the squared speed of sound is very close to $1/3$ times the speed of light squared, and~there is an increase in the value with increasing $\big<\diff{N_{ch}}\eta\big>$, suggesting that the system becomes more ideal at larger multiplicity. This observation complements the near-ideal behaviour already indicated from the measurement of isothermal~compressibility.
 
\section{Conclusions} \label{sec:Summary}

With an aim towards understanding the system produced in the heavy-ion collision, we have made an attempt to study some thermodynamic response functions such as isothermal compressibility and speed of sound. Since transverse momentum spectra carries information about the system, we have analyzed spectra of charged hadrons at three different LHC energies using the unified formalism and used the extracted value of thermodynamical parameters to study the isothermal compressibility and speed of sound. The~\pte-spectra of charged hadrons produced in \pb collision at $2.76$ TeV, $5.02$ TeV and $Xe-Xe$ collision at 5.44 TeV are taken with \pts range upto 5 $GeV/c$. We have estimated the value of $\kappa_T/V$ and $c_s^2$ and studied their variation as a function of charged particle multiplicity. We observed that while the value of  $\kappa_T/V$ decreases with respect to increase in multiplicity, the~values of $c_s^2$ approaches to 1/3.

These estimations of $\kappa_T/V$ and $c_s^2$ using unified formalism represent that the medium tends to move toward a near-ideal behavior with an increase in charged particle multiplicity. In~conclusion, we have presented the theoretical formalism to study some of the thermodynamical response functions within the unified statistical framework discussed in the Ref.~\cite{Gupta:2020naz}. The~extracted values point toward the creation of a near-ideal medium in high-energy collision and the system approach the ideal behavior as we move from peripheral to the central~collision.

\section{Acknowledgement} 
 We would like to acknowledge that the work has been carried out using the computing facility in EHEP lab in Department of Physics at IISER Mohali.

\end{document}